\documentclass[aoas,preprint]{imsart}

\RequirePackage[OT1]{fontenc}
\RequirePackage{amsthm,amsmath,amssymb}
\usepackage{amsfonts,mathrsfs,stmaryrd}
\RequirePackage[numbers]{natbib}
\RequirePackage[colorlinks,citecolor=blue,urlcolor=blue]{hyperref}
\usepackage{graphicx}

\usepackage{array}
\usepackage{multirow}

\startlocaldefs
\numberwithin{equation}{section}
\theoremstyle{plain}
\newtheorem{thm}{\hspace*{1pc}Theorem}
\newtheorem{proposition}{\hspace*{1pc}Proposition}
\newtheorem{corollary}{\hspace*{1pc}Corollary}

\endlocaldefs

\newcommand{\cL}{\mathcal{L}}

\newcommand{\sN}{\mathscr{N}}

\newcommand{\sS}{\mathscr{S}}

\newcommand{\mP}{\mathbb{P}}

\begin{document}

\begin{frontmatter}
\title{Fast Parameter Estimation in Loss Tomography for Networks of General Topology\thanksref{T1}}
\runtitle{Loss Tomography for Network of General Topology}
\thankstext{T1}{This study is partially supported by National Science Foundation of USA grant DMS-1208771,  National Science Foundation of China grant 11401338, Shenzhen Fund for Basic Science grant JC201005280651A and grant JC201105201150A.}

\begin{aug}
\author{\fnms{Ke} \snm{Deng}\thanksref{t1}\ead[label=e1]{kdeng@math.tsinghua.edu.cn}},
\author{\fnms{Yang} \snm{Li}\thanksref{t3}\ead[label=e2]{yli01@fas.harvard.edu}},
\author{\fnms{Weiping} \snm{Zhu}\thanksref{t2}\ead[label=e3]{w.zhu@adfa.edu.au}}
\and
\author{\fnms{Jun S.} \snm{Liu}\thanksref{t3}
\ead[label=e4]{jliu@stat.harvard.edu}
\ead[label=u1,url]{http://www.people.fas.harvard.edu/~junliu/}}

\runauthor{K. Deng et al.}

\affiliation{Tsinghua University\thanksmark{t1}, University of New South Wales\thanksmark{t2} and Harvard University\thanksmark{t3}}

\address{Address of the First author\\
Yau Mathematical Sciences Center \& Center for Statistical Science,
Tsinghua University, Beijing 100084, China\\
\printead{e1}}

\address{Address of the Second author\\
Department of Statistics, Harvard University, Cambridge, MA 02138, USA\\
\printead{e2}}

\address{Address of the Third author\\
Department of ITEE, ADFA@University of New South Wales, Canberra ACT 2600, Australia\\
\printead{e3}}

\address{Address of the Fourth author\\
Department of Statistics, Harvard University, Cambridge, MA 02138, USA\\
\printead{e4}\\
}
\end{aug}

\begin{abstract}
As a technique to investigate link-level loss rates of a computer network with low operational cost, loss tomography has received considerable attentions in recent years. A number of parameter estimation methods have been proposed for loss tomography of networks with a tree structure as well as a general topological structure. However, these methods suffer from either high computational cost or insufficient use of information in the data. In this paper, we provide both theoretical results and practical algorithms for parameter estimation in loss tomography. By introducing a group of novel statistics and alternative parameter systems, we find that the likelihood function of the observed data from loss tomography keeps exactly the same mathematical formulation for tree and general topologies, revealing that networks with different topologies share the same mathematical nature for loss tomography. More importantly, we discover that a re-parametrization of the likelihood function belongs to the standard exponential family, which is convex and has a unique mode under regularity conditions. Based on these theoretical results, novel algorithms to find the MLE are developed. Compared to existing methods in the literature, the proposed methods enjoy great computational advantages.
\end{abstract}


\begin{keyword}
\kwd{network tomography}
\kwd{loss tomography}
\kwd{general topology}
\kwd{likelihood equation}
\kwd{pattern-collapsed EM algorithm}
\end{keyword}

\end{frontmatter}

\section{Introduction}\label{sec:Introduction}

Network characteristics such as loss rate, delay, available
bandwidth, and their distributions are critical to various network
operations and important for understanding network behaviors. Although
considerable attention has been given to network measurements, due to
various reasons (e.g., security, commercial interest and
administrative boundary) some  characteristics of the network cannot be
obtained directly from a large network. To overcome this difficulty,
network tomography was proposed in \cite{Vardi1996}, suggesting the
use of end-to-end measurement and statistical inference to estimate
characteristics of a large network. In an end-to-end measurement, a
number of sources are attached to the network of interest to send
probes to receivers attached to the other side of the network,
and  paths from  sources to receivers cover links of
interest. Arrival orders and arrival times of the probes
carry the information of the network, from which many network
characteristics can be inferred statistically. 
Characteristics that have been estimated in this manner include
link-level loss rates
\cite{CDHT99,CDMT99,CDMT99a,HBB00,CN00,CN001,BDPT,Duffield2002,DHTWF02,ZG04-3,ZG05,DHPT06},
delay distributions
\cite{DHPT01,Presti2002,LY03,TCN03,SH03,Lawrence06,Arya2008,Dinwoodie2007,Chen2010,Deng2012},
origin-destination traffic
\cite{Vardi1996,LY03,Bell1991,de2011modelling,cascetta1988unified,yang1992estimation,lo1996estimation,bianco2001network,Tebaldi1998,Cao2001,Cao2002,Medina2002,Zhang2003,Castro2004,LTY06,Fang2007,Airoldi2011},
loss patterns \cite{ADV07}, and the network topology \cite{RNC03}. In this
paper, we focus on the problem of estimating loss rates.

Network topologies connecting sources to receivers can be divided into
two classes: tree and general. A tree topology, as named, has a
single source attached to the root of a multicast tree to send
probes to receivers attached to the leaf nodes of the tree. A
network with a  general topology, however, requires a number of trees
to cover all links of the network. Each tree has one source sending
probes to receivers in it. Because of the use of multiple sources to
send probes in a network with general topology, those nodes and receivers located at
intersections of multiple trees can receive probes from multiple
sources. In this case, we must consider impacts of probes sent
by all sources simultaneously in order to get a good estimate.
This task is much more challenging than the tree topology case.

Numerous  methods have been proposed for loss tomography in a tree
topology. C$\acute{a}$ceres \emph{et al.} \cite{CDHT99} used a
Bernoulli model to describe the loss behavior of a link, and
derived the MLE for the pass rate of a path connecting the source to
a node, which was expressed as the solution of a polynomial equation
\cite{CDHT99,CDMT99,CDMT99a}. To ease the concern of using numerical
methods to solve a high degree polynomial, several papers have been
published to accelerate the calculation at the price of a little accuracy loss: 
 Zhu and Geng proposed a recursively defined
estimator based on a bottom-up strategy in \cite{ZG04-3,ZG05};
Duffield \emph{et al.} proposed a closed-form estimator in
\cite{DHPT06}, which has the same asymptotic variance as the MLE to
the first order. Considering the unavailability of multicast in some
networks, Harfoush {\it et al.} \cite{HBB00} and Coates {\it et al.}
\cite{CN00} independently proposed the use of the unicast-based
multicast to send probes to receivers, where Coates {\it et al.}
also suggested the use of the EM algorithm \cite{Dempster1977} to
estimate link-level loss rates.

For networks beyond a tree, however, little research has been done, although a majority of networks in practice fall into this category. Conceptually, the topology of a general network can be arbitrarily complicated. However, no matter how complicated a general topology is, it can always be covered by a group of carefully selected trees, in each of which an end-to-end experiment can be carried out independently to study the properties of the subnetwork covered by the tree. If these trees do not overlap with each other, the problem of studying the whole general network can be decomposed into a group of smaller subproblems, each for one tree. However, due to the complexity of the network topology and practical constraints, it is more than often that the selected trees overlaps significantly, i.e., some links are shared by two or more trees (see Figure \ref{L5net} for example). The shared links of two selected trees induce dependence between the two trees. Simply ignoring the dependence leads to a loss of information. How to effectively integrate the information from multiple trees to achieve a joint analysis is a major challenge in network tomography of general topology.

The first effort on network tomography of general topology is due to Bu {\it et al.} \cite{BDPT}, who attempted to extend the method in \cite{CDHT99} to networks with general topology. Unfortunately, the authors failed to derive an explicit expression for the MLE like the one presented in \cite{CDHT99} for this more general case. They then resorted to an iterative procedure (i.e., the EM algorithm) to search for the MLE. In addition, a heuristic method, called \emph{minimum variance weighted average} (MVWA) algorithm, was also proposed in \cite{BDPT}, which deals with each tree in a general topology separately and averages the results. The MVWA algorithm is less efficient than the EM algorithm, especially when the sample size is small. Rabbat {\it et al.} in \cite{RNC03} considered the tomography problem for networks with an unknown but general topology, mainly focusing on network topology identification, which is beyond the scope of our current paper. 

In this paper, we provide a new perspective for the study of loss tomography, which is applicable to both tree and general topologies. Our theoretical contributions are: 1) introducing a set of novel statistics, which are complete and minimal sufficient; 2) deriving two alternative parameter systems and the corresponding re-parameterized likelihood functions, which benefit us both theoretically and computationally; 3) discovering that the loss tomography for a general topology shares the same mathematical formulation as that of a tree topology; and, 4) showing that the likelihood function belongs to the exponential family and has a unique mode (which is the MLE) under regularity conditions. Based on these theoretical results, we propose two new algorithms (a likelihood-equation-based algorithm called LE-$\xi$ and an EM-based algorithm called PCEM) to find MLE. Compared to existing methods in the literature, the proposed methods are computationally much more efficient.

The rest of the paper is organized as follows. Section 2 introduces notations for tree topologies and the stochastic model for loss rate inference. Section 3 describes a set of novel statistics and two alternative parameter systems for loss rate analysis in tree topologies. New forms of the likelihood function are established based on the novel statistics and re-parametrization. Section 4 extends the above results to general topologies. Section 5 and 6 propose two new algorithms for finding MLE of $\theta$, which enjoy great computational advantages over existing methods. Section 7 evaluates performances of the proposed methods by simulations. We conclude the article in Section 8.

\section{Notations and Assumptions}\label{sec:NotationAssumption}

\subsection{Notations for Tree Topologies}
\begin{figure}
\setlength{\unitlength}{1cm}
\begin{picture}(4,3.5)
\put(0.3,-0.2){\includegraphics[scale=0.5]{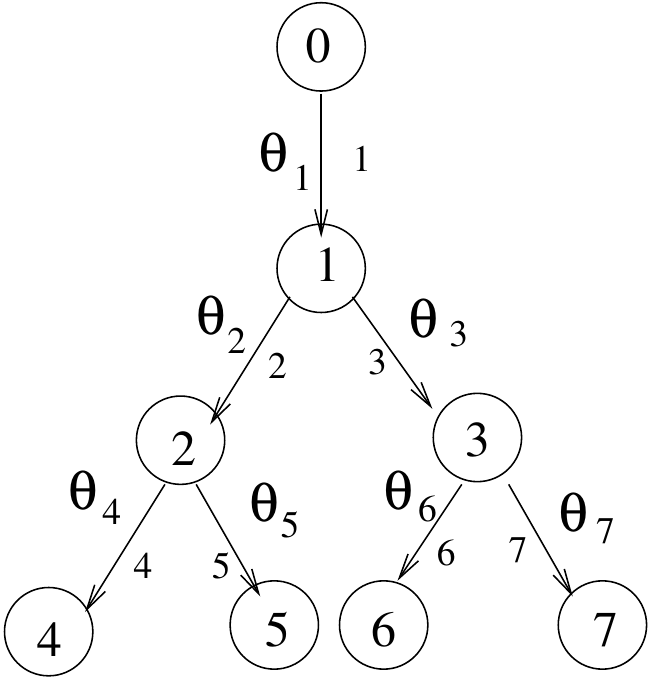}}
\end{picture}
\caption{A multicast tree.} \label{struct}
\end{figure}

We use $T=(V, E)$ to denote the multicast tree of
interest, where $V=\{v_0, v_1, ... v_m\}$ is a set of nodes
representing the routers and switches in the network, and
$E=\{e_1,..., e_m\}$ is a set of directed links connecting the
nodes. Two nodes connected by a directed link are called the \emph{parent
node} and the \emph{child node}, respectively, and the direction of
the arrow indicates that  the parent 
forwards received probes to the child. Figure \ref{struct} shows a
typical multicast tree. Note that the root node of a
multicast tree has only one child, which is slightly different from
an ordinary tree, and each non-root node has exactly one parent.

Each link is assigned a unique ID number ranging from $1$ to $m$, based on which each node  
obtains its unique ID number ranging from $0$ to $m$ correspondingly so that
link $i$ connects node $i$ with its parent node. Number $0$ is
reserved for the source node. In contrast to \cite{CDHT99} and
\cite{DHPT06}, whose methods are node-centric, our methods here
focus on links instead. For a network with tree topology, the two reference systems are equivalent as there exists an one-to-one mapping between nodes and links: every node in the network (except for the source) has a unique parent link. For a network with general topology, however, the link-centric system is more convenient, as a node in the network may have multiple parent links.

We let $f_i$ denote the unique parent link, $B_i$ the brother links, and $C_i$ the child links, respectively, of link $i$. To be precise, $B_i$ contains all links that share the same parent node with link $i$, including link $i$ itself. A subtree $T_i=\{V_i, E_i\}$ is defined as the subnetwork composed of link $i$ and all its descendant links. We let $R$ and $R_i$ denote the set of receivers (i.e., leaf nodes) in $T$ and $T_i$, respectively. Sometimes, we also use $R$ or $R_i$ to denote the leaf links of $T$ or $T_i$. The concrete meaning of $R$ and $R_i$ can be determined based on the context.

Taking the multicast tree displayed in Figure \ref{struct} as an example, we have $V=\{0,1,\cdots,7\}$, $E=\{1,2,\cdots,7\}$ and $R=\{4,5,6,7\}$. For the subtree $T_2$, however, we have $V_2=\{1,2,4,5\}$, $E_2=\{2,4,5\}$ and $R_2=\{4,5\}$. For link 2, its parent link is $f_2=\{1\}$, its brother links are $B_2=\{2,3\}$, and its child links are $C_2=\{4,5\}$.

\subsection{Stochastic Model}
In a multicast experiment, $n$ probe packages are sent from the root
node $0$ to all the receivers. Let $X_i^{(t)}=1$ if the $t$-th probe
package reached node $i$, and $0$ otherwise. The status of probes at
receivers $\{X_r^{(t)}\}_{r\in R,1\leq t\leq n}$ can be directly
observed from the multicast experiment; but, the status of probes at
internal links cannot be directly observed. In the following, we use
$\textbf{X}_R=\{X^{(t)}_R\}_{1\leq t\leq n}$, where $X^{(t)}_R=\{X^{(t)}_r\}_{r\in R}$, to denote the data collected in a multicast experiment in tree $T$.

In this paper, we model the loss behavior of links by the
Bernoulli distribution and assume the \emph{spatial-temporal
independence} and \emph{temporal homogeneity} for the network, i.e.
the event $\{X_i^{(t)}=0\mid X_{f_i}^{(t)}=1\}$ are independent
across $i$ and $t$, and
$$P(X_i^{(t)}=0\mid X_{f_i}^{(t)}=0)=1,\ P(X_i^{(t)}=0\mid X_{f_i}^{(t)}=1)=\theta_i.$$
We call $\theta=\{\theta_i\}_{i\in E}$ the \emph{link-level loss
rates}, and the goal of loss tomography is to estimate $\theta$ from
$\textbf{X}_R$.

\subsection{Parameter Space}
In principle, $\theta_i$ can be any value in $[0,1]$. Thus, the natural parameter space of $\theta$ is $\Theta^*=[0,1]^m$, an $m$-dimensional closed unit cube. In this paper, however, we assume that $\theta_i\in(0,1)$ for every $i\in E$, and thus constrain the parameter space to $\Theta=(0,1)^m$, to simplify the problem. If $\theta_i=1$ for some $i\in E$, then the subtree $T_i$ is actually disconnected from the other part of the network, since no probes can go through link $i$. In this case, the loss rate of other links in $T_i$ are not estimable due to the lack of information. On the other hand, if $\theta_i=0$ for some $i\in E$, the original network of interest degenerates to an equivalent network where node $i$ is removed and all its child nodes are connected directly to node $i$'s parent node. By constraining the parameter space of $\theta$ into $\Theta$, we exclude these degenerate cases from consideration.

\section{Statistics and Likelihood Function}\label{sec:StatisticsLikelihood}

\subsection{The likelihood function and the MLE}
Given the loss model for each link, we can write down the likelihood function and use the maximum likelihood principle to determine unknown parameters.  That is, we aim to find the parameter value that maximizes the log-likelihood function:
\begin{eqnarray}\label{obj}
\arg\max_{\theta\in\Theta}L(\theta)=\arg\max_{\theta\in\Theta}\sum_{x \in \Omega}n(x)\log P(x; \theta), 
\end{eqnarray}
where $x$ stands for the observation at receivers (i.e., a realization of $X_R$), $\Omega=\{0,1\}^{|R|}$ is the space of all possible observations with $|R|$ denoting the size of set $R$, $n(x)$ is the number of occurrences of observation $x$, and
$P(x;\theta)$ is the probability of observing $x$ given  parameter value
$\theta$. However, the log-likelihood function (\ref{obj}) is more
symbolic than practical because of the following reasons: (1) evaluating the log-likelihood function (\ref{obj}) is an expensive operation as it needs to scan through all possible $x\in\Omega$; and, (2) the likelihood equation derived from (\ref{obj}) cannot be solved analytically, and it is often computationally expensive to pursue a numerical solution.

\subsection{Internal State and Internal View}\label{internal view1}
Instead of using the log-likelihood function (\ref{obj}) directly,
we consider to rewrite it in a different form. Under the posited probabilistic
model, the overall likelihood $P(\textbf{X}_R\mid
\theta)$ is the product of the likelihood from each probe, i.e.,
$$P(\textbf{X}_R\mid \theta)=\prod_{t=1}^n P(X_R^{(t)}\mid\theta).$$
Thus, an alternative form of the overall likelihood can be obtained
by explicating the likelihood of each single probe and accumulating
them. Two concepts called \emph{internal state} and \emph{internal
view} can be generated from this process.

\subsubsection{Internal State}
For a link $i\in E$, given the observation of probe $t$ at $R_i$
and $R_{f_i}$, we are able to partially confirm whether the probe
passes link $i$. Formally, for observation $\{X_j^{(t)}\}_{j\in
R_i}$, we define $Y_i^{(t)}=\max_{j \in R_i} X_j^{(t)}$ as the
\emph{internal state} of link $i$ for probe $t$.
If $Y_i^{(t)}=1$, probe $t$ reaches at least one receiver attached
to $T_i$, which  implies that the probe passes link $i$.
Furthermore, by considering  $Y_{f_i}^{(t)}$ and $Y_i^{(t)}$ simultaneously,
we have three possible scenarios for each internal node $i$:
\begin{itemize}
\item  $Y_{f_i}^{(t)}=Y_i^{(t)}=1$, i.e., we observed that probe $t$ passed link $i$; or
\item  $Y_{f_i}^{(t)}=1$ and $Y_i^{(t)}=0$, i.e., we observed that probe $t$ reached node
$f_i$, but we did not know whether it reached node $i$ or not; or
\item $Y_{f_i}^{(t)}=Y_i^{(t)}=0$, i.e., we did not know whether probe $t$ reached node $f_i$ or not at all;
\end{itemize}
as  $Y_{f_i}^{(t)}=0$ and $Y_i^{(t)}=1$ can never happen by definition of $Y$.

The three scenarios have different impacts on the likelihood
function. Formally, define
\begin{eqnarray*}
E_{t,1}&=&\{i\in E:Y_i^{(t)}=1\},\\
E_{t,2}&=&\{i\in E:Y_{f_i}^{(t)}=1, Y_i^{(t)}=0\},\\
E_{t,3}&=&\{i\in E:Y_{f_i}^{(t)}=Y_i^{(t)}=0\}.
\end{eqnarray*}
We have
\begin{equation}\label{LikelihoodOfSingleProbe}
P(X_R^{(t)}\mid\theta)=\prod_{i\in E_{t,1}}(1-\theta_i)\prod_{i\in
E_{t,2}}\xi_i(\theta),
\end{equation}
where
$$\xi_i(\theta)=P(X_j=0,\ \forall\ j\in R_i\mid
X_{f_i}=1;\theta)$$
represents 
the probability that a probe sending out from the root node of $T_i$
fails to reach any leaf node in $T_i$.

\subsubsection{Internal View}
Accumulating the internal states of each link in the experiment, we
have
\begin{equation}\label{InternalView}
n_i(1)=\sum_{t=1}^n Y_i^{(t)},
\end{equation}
which counts the number of probes whose pass through link $i$ can be confirmed
from observations. Specifically, we  define $n_0(1)=n$. Moreover, define
\begin{equation}\label{InternalView2}
n_i(0)=n_{f_i}(1)-n_i(1),
\end{equation}
for $\forall\ i\in E$. We call
the statistics $\{n_i(1),n_i(0)\}$ the \emph{internal view} of link
$i$. Based on internal views, we can write the log-likelihood
of $\textbf{X}_R$ in a more convenient form:
$$L(\theta)=\sum_{i\in E}\Big[n_i(1)\log(1-\theta_i)+n_i(0)\log\xi_i(\theta)\Big].$$

\subsection{Re-parametrization}
Two alternative parameter systems can be introduced to re-parameterize the above log-likelihood function. First, based on the definition of
$\xi_i(\theta)$, we have
\begin{equation}\label{theta to xi}
\xi_i(\theta)=\theta_i+(1-\theta_i)\prod_{j \in C_i} \xi_j(\theta),\
\ i\in E.
\end{equation}
Note that if $i\in R$, we have $C_i=\emptyset$, and
(\ref{theta to xi}) degenerates to $\xi_i(\theta)=\theta_i$. Let
$\xi_i\triangleq\xi_i(\theta)$ for $i\in E$. Equation (\ref{theta to
xi}) defines a one-to-one mapping between two parameter
systems $\theta=\{\theta_i\}_{i\in E}$ and $\xi=\{\xi_i\}_{i\in E}$,
i.e.,
$$\Gamma:\ \Theta\mapsto\Xi,\ \xi=\Gamma(\theta)\triangleq\big(\xi_1(\theta),\cdots,\xi_m(\theta)\big),$$
where $\Theta$ and $\Xi$ are the domain and image, respectively. The inverse mapping of $\Gamma$ is
$$\Gamma^{-1}:\ \Xi\mapsto\Theta,\ \theta=\Gamma^{-1}(\xi)\triangleq\big(\theta_1(\xi),\cdots,\theta_m(\xi)\big),\ \mbox{where}$$
\begin{equation}\label{xi to theta}
\theta_{i}(\xi)=\frac{\xi_{i}-\prod_{j\in C_{i}}\xi_j}{1-\prod_{j\in
C_{i}}\xi_j},\ \ i\in E.
\end{equation}
Using $\xi$ to replace $\theta$ in $L(\theta)$, we
have the following alternative log-likelihood function with $\xi$ as
parameters:
$$L(\xi)=\sum_{i\in E}\Big[{n_i(1)}\log(\frac{1-\xi_i}{1-\prod_{j\in C_{i}}\xi_j})+{n_i(0)\log \xi_i}\Big].$$

Second, $L(\xi)$ can be further re-organized into
\begin{eqnarray*}
L(\xi)&=&\sum_{i\in E}n_i(1)\log(\frac{1-\theta_i(\xi)}{\xi_i})+\sum_{i\in E}n_{f_i}(1)\log\xi_i\\
&=&n\log\xi_1+\sum_{i\in E}n_i(1)\log\psi_i(\xi),
\end{eqnarray*}
where $\psi_i(\xi)$ is defined as
\begin{equation}\label{xi to psi}
\psi_{i}(\xi)=\left\{
  \begin{array}{ll}
  \log\frac{1-\theta_i(\xi)}{\xi_i},&i\in R,\\
  \log\frac{\xi_i-\theta_i(\xi)}{\xi_i},&i\notin R.
  \end{array}
\right.
\end{equation}
Similarly, let $\psi_i\triangleq\psi(\xi)$ for $i\in E$. Equation
(\ref{xi to psi}) defines a one-to-one mapping between parameter
system $\xi=\{\xi_i\}_{i\in E}$ and $\psi=\{\psi_i\}_{i\in E}$,
i.e.,
$$\Lambda:\ \Xi\mapsto\Psi,\ \psi=\Lambda(\xi)\triangleq\big(\psi_1(\xi),\cdots,\psi_m(\xi)\big),$$
where $\Psi\triangleq\Lambda(\Xi)$ is the image of $\psi$. The
inverse mapping of $\Lambda$ is
$$\Lambda^{-1}:\ \Psi\mapsto\Xi,\ \xi=\Lambda^{-1}(\psi)\triangleq\big(\xi_1(\psi),\cdots,\xi_m(\psi)\big).$$
It can be shown that
$$\psi_i=\log P\big(X_i=1\mid X_{f_i}=1;\ X_j=0,\ \forall\ j\in R_i\big)\ \mbox{for}\ i\notin R,$$
from which the physical meaning of $\psi_i$ can be better understood. Using $\psi$ to
replace $\xi$ in $L(\xi)$, we have the following
log-likelihood function with $\psi$ as parameters:
$$L(\psi)=n\log \xi_1(\psi)+\sum_{i\in E}n_i(1)\psi_{i}.$$

To illustrate the relations of $\theta$, $\xi$ and $\psi$, let's consider the toy network in Figure \ref{struct} with $\theta_i=0.1$ for $i=1,\cdots,7$. It is easy to check that:
\begin{eqnarray*}
\xi_4=\xi_5=\xi_6=\xi_7&=&0.1,\\
\xi_2=\xi_3&=&0.1+(1-0.1)\times0.1^2=0.109,\\
\xi_1&=&0.1+(1-0.1)\times0.109^2\approx0.1107;\\
\psi_4=\psi_5=\psi_6=\psi_7&=&\log\frac{1-0.1}{0.1}\approx2.1972,\\
\psi_2=\psi_3&=&\log\frac{0.109-0.1}{0.109}\approx-2.4941,\\
\psi_1&\approx&\log\frac{0.1107-0.1}{0.1107}\approx-2.3366.
\end{eqnarray*}
We list these concrete values in Table \ref{ToyExample} for comparison purpose. The notations, statistics and parameter systems are summarized into Table \ref{symbols} for easy reference.

\begin{table}[h]
\caption{Comparing different parameter systems for the toy network in Figure \ref{struct}} \label{ToyExample}
\begin{center}
    \begin{tabular}{c ccccccc}
    \hline
Link	&	1	&	2	&	3	&	4	&	5	&	6	&	7	\\\hline
$\theta$	&	0.1	&	0.1	&	0.1	&	0.1	&	0.1	&	0.1	&	0.1	\\
$\xi$	&	$0.1092$	&	$0.109$	&	$0.109$	&	0.1	&	0.1	&	0.1	&	0.1	\\
$\psi$	& $-2.3366$		& $-2.4941$	&	$-2.4941$	&	$2.1972$	&	$2.1972$	&	$2.1972$	&	$2.1972$	\\\hline
\end{tabular}
\end{center}
\end{table}

\begin{table}[h]
\caption{Collection of symbols} \label{symbols}
\begin{center}
    \begin{tabular}{|c|l|}
    \hline
    {\bf Symbol} & $\hspace{2.8cm}${\bf Meaning }\\ \hline
    $T$     & the multicast tree of interest  \\ \hline
    $V$     & the node set of $T$  \\ \hline
    $E$     & the link set of $T$  \\ \hline
    $R$     & the set of leaf nodes (receivers) or leaf links of $T$ \\ \hline
    $T_i$   & the subtree with link $i$ as the root link\\ \hline
    $V_i$   & the node set of subtree $T_i$ \\ \hline
    $R_i$   & the set of leaf nodes (receivers) or leaf links of $T_i$ \\ \hline
    $f_i$   & the parent link of link $i$ \\\hline
    $B_i$   & the brother links of link $i$ \\ \hline
    $C_i$   & the child links of link $i$ \\ \hline
    $X_i^{(t)}$ & $X_i^{(t)}=1$ if probe $t$ reached node $i$, and $0$ otherwise\\ \hline
    $Y_i^{(t)}$ & $\max_{j\in R_i}X_j^{(t)}$\\ \hline
    $n_i(1)$ & $\sum_{t=1}^nY_i^{(t)}$, number of probes that passed link $i$ for sure\\ \hline
    $n_i(0)$ & $n_{f_i}(1)-n_i(1)$\\ \hline
    $\theta_i$& $P(X_i=0\mid X_{f_i}=1)$, the loss rate of link $i$ \\ \hline
    $\xi_i$& $P(X_j=0,\ \forall\ j\in R_i\mid X_{f_i}=1)$, the loss rate of $T_i$ \\ \hline
    $\psi_i$& $\log P(X_i=1\mid X_{f_i}=1; X_j=0,\ \forall\ j\in R_i)$ \\ \hline
    \end{tabular}
\end{center}
\end{table}

\section{Likelihood Function for General Networks}\label{sec:LikelihhodForGeneralNet}


\begin{figure}
\setlength{\unitlength}{1cm}
\includegraphics[scale=0.425]{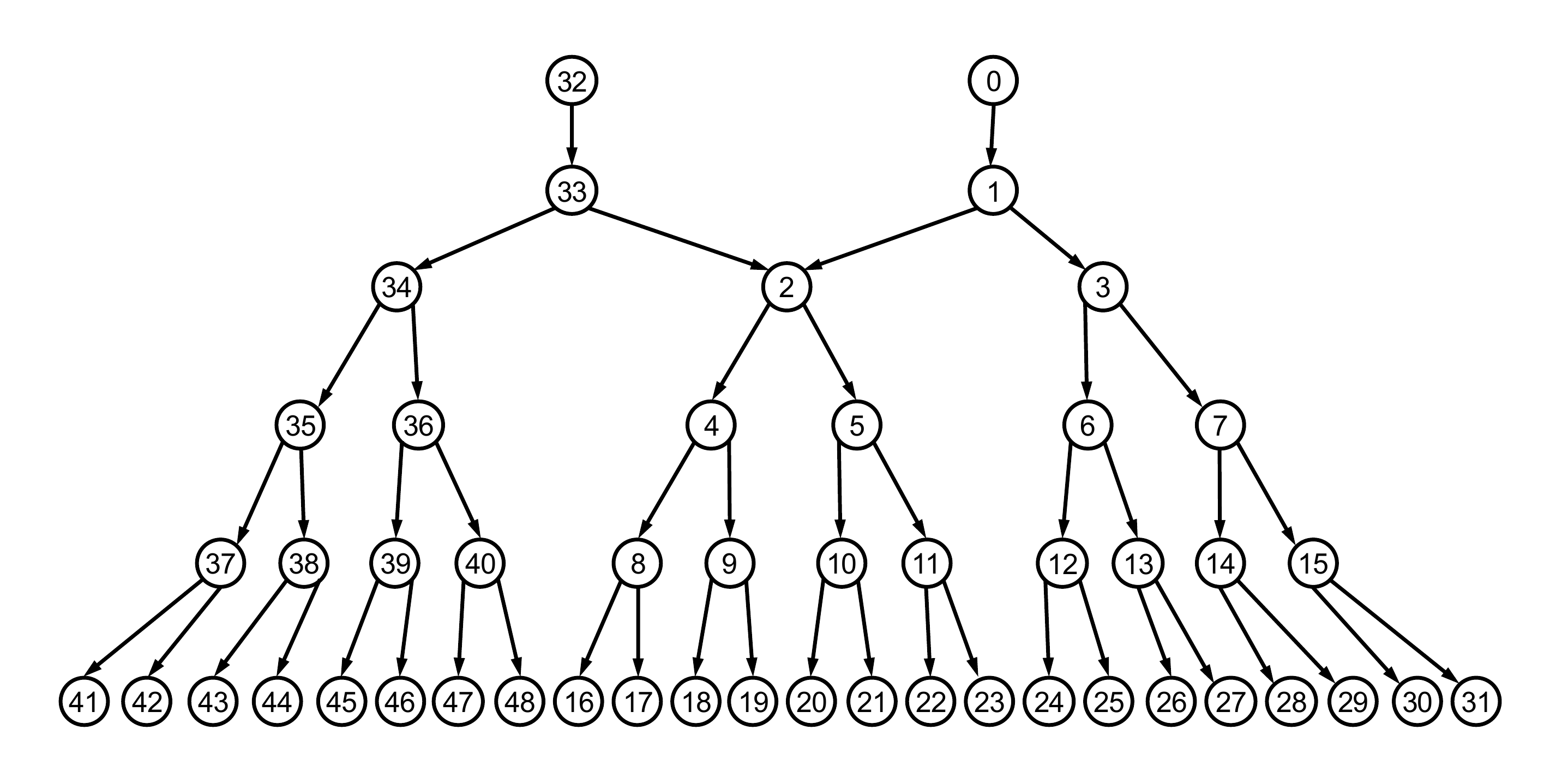}
\caption{A 5-layer network covered by two trees.}
\label{L5net}
\end{figure}

In this section, we will extend the concept of internal view and alternative parameter systems to general networks containing multiple trees. Formally, we use $\sN=\{T_1,\cdots,T_K\}$ to denote a general network covered by $K$ multicast trees, where each multicast tree $T_k$ covers a subnetwork with $S_k$ as the root link. For example, Figure \ref{L5net} illustrates a network with $K=2$, $S_1=0$ and $S_2=32$.  Let $\sS=\{S_1,\cdots,S_K\}$ be the set of root links in $\sN$.

For each tree $T_k$, let $\{n_{k,i}(1),n_{k,i}(0)\}$ be the internal view of link
$i\in E_k$ in $T_k$ based on the $n_k$ probes sent out from $S_k$.
Specially, define $n_{k,i}(1)=0$ for link $i\notin E_k$. The experiment on $T_k$ results in the following log-likelihood functions with $\theta$, $\xi$ and $\psi$ as parameters, respectively:
\begin{eqnarray*}
L_k(\theta)&=&\sum_{i\in E}\Big[n_{k,i}(1)\log(1-\theta_i)+n_{k,i}(0)\log\xi_i(\theta)\Big],\\
L_k(\xi)&=&\sum_{i\in E}\Big[{n_{k,i}(1)}\log(\frac{1-\xi_i}{1-\prod_{j\in C_{i}}\xi_j})+{n_{k,i}(0)\log \xi_i}\Big],\\
L_k(\psi)&=&n_k\log \xi_{S_k}(\psi)+\sum_{i\in E}n_{k,i}(1)\psi_{i}.
\end{eqnarray*}

Considering that a multicast experiment in a general network $\sN$ is
a pool of $K$ independent experiments in the $K$ trees
of $\sN$, the log-likelihood function of the whole experiment is just the summation of $K$ components, i.e.,
$$L(\theta)=\sum_{k=1}^K L_k(\theta),\ \ L(\xi)=\sum_{k=1}^K L_k(\xi)\ \ \mbox{and}\ \ L(\psi)=\sum_{k=1}^K L_k(\psi).$$
Define the internal view of link $i$ in a general network as:
\begin{equation}\label{InternalViewGeneral}
n_i(1)=\sum_{k=1}^Kn_{k,i}(1),\ \ n_{i}(0)=\sum_{k=1}^Kn_{k,i}(0).
\end{equation}
It is straightforward to see that
\begin{eqnarray}
L(\theta)&=&\sum_{i\in E}\Big[n_i(1)\log(1-\theta_i)+n_i(0)\log\xi_i(\theta)\Big],\\
\label{LogLikelihoodGeneralXi}
L(\xi)&=&\sum_{i\in E}\Big[{n_{i}(1)}\log(\frac{1-\xi_i}{1-\prod_{j\in C_{i}}\xi_j})+n_{i}(0)\log \xi_i\Big],\\
\label{LogLikelihoodGeneralPsi}
L(\psi)&=&\sum_{k=1}^Kn_k\log\xi_{S_k}(\psi)+\sum_{i\in E}n_i(1)\psi_{i}.
\end{eqnarray}
Note that in this general case, we have:
\begin{eqnarray*}
 && \mbox{for}\ i\in\sS,\ n_{S_k}(1)=n_k;\ \mbox{and}\\
 && \mbox{for}\ i\notin\sS,\ n_{i}(0)+n_i(1)=\sum_{j\in F_i}n_j(1).
\end{eqnarray*}
Here, $F_i$ stands for the unique or multiple parent links of link $i$ in the general network $\sN$.

\section{Likelihood Equation}\label{sec:LikelihoodEquation}
The bijections among the the three parameter systems mean that we can switch among them freely without changing the result of parameter estimation:
\begin{proposition}\label{LinkTheoremAmongParameters}
The results of likelihood inference based on the three parameter
systems are equivalent, i.e.,
$$\Gamma\big(\arg\max_{\theta\in\Theta}L(\theta)\big)=\arg\max_{\xi\in\Xi}L(\xi)=\Lambda^{-1}\big(\arg\max_{\psi\in\Psi}L(\psi)\big);$$
and the likelihood equations under different parameters share the
same solution, i.e.,
$$\frac{\partial L(\theta)}{\partial\theta}\Big|_{\theta=\theta^*}=0\Leftrightarrow
\frac{\partial
L(\xi)}{\partial\xi}\Big|_{\xi=\xi^*}=0\Leftrightarrow
\frac{\partial L(\psi)}{\partial\psi}\Big|_{\psi=\psi^*}=0,$$ if
$\theta^*\in\Theta$, or $\xi^*\in\Xi$, or $\psi^*\in\Psi$, where
$\Gamma(\theta^*)=\xi^*=\Lambda^{-1}(\psi^*)$.
\end{proposition}

This flexibility provides us great theoretical and computational advantages. On the the theoretical aspect, as $\{n_k\}_{k=1}^K$ are known constants,  $L(\psi)$ falls into the standard exponential family with $\psi$ as the natural parameters. Thus, based on the properties of the exponential family \cite{LC98}, we have the following results immediately:
\begin{proposition}\label{SuffStat}
The following results hold for loss tomography:
\begin{enumerate}
  \item Statistics $\{n_i(1)\}_{i\in E}$ are complete and minimal
sufficient;
  \item The likelihood equation $\frac{\partial
L(\psi)}{\partial\psi}=0$ has at most one solution $\psi^*\in\Psi$;
  \item If $\psi^*$ exists, $\psi^*$ (or $\theta^*=(\Lambda\circ\Gamma)^{-1}(\psi^*)$) is the MLE.
\end{enumerate}
\end{proposition}

On the computational aspect, the parameter system $\xi$ plays a central role. Different from the likelihood equation with $\theta$ as parameters, which is intractable, the likelihood equation with $\xi$ as parameters enjoys unique computational advantages. Let
$$r_i=\frac{n_i(1)}{n_i(1)+n_i(0)}.$$
It can be shown that the likelihood equation with $\xi$ as parameters is:
\begin{equation}\label{LikelihoodEquationXi}
\xi_i=(1-r_i)+r_i\cdot\prod_{j\in B_i}\xi_j\cdot I(i\notin\sS),
\end{equation}
for any link $i\in E$.

For link $i\in\sS$, (\ref{LikelihoodEquationXi}) degenerates to
$$\xi_i=1-r_i.$$
For link $i\notin\sS$, define $\pi_i=\prod_{j\in B_i} \xi_j$. We note that solving $\xi_i$ from (\ref{LikelihoodEquationXi}) is equivalent to solving the following equation about $\pi_i$:
\begin{equation}\label{LikelihoodEquationPi}
\pi_i=\prod_{j\in B_i}\big[(1-r_j)+r_j \cdot\pi_i\big],
\end{equation}
which can be solved analytically when $|B_i|=2$, and by numerical approaches \cite{Todd76} when $|B_i|>2$.

From (\ref{LikelihoodEquationXi}) and (\ref{LikelihoodEquationPi}), it is transparent that only local statistics $\{r_j\}_{j\in B_i}$ are involved in estimating $\xi_i$. This observation leads to the following fact immediately: if a processor keeps the values of $\{r_j\}_{j\in B_i}$ in its memory, $\{\xi_j\}_{j\in B_i}$ can be effectively estimated by the processor independent of estimating the other parameters. Based on this unique property of the likelihood equation with $\xi$ as parameter, we propose a parallel procedure called ``LE-$\xi$" algorithm to estimate $\theta$ (see Algorithm 1).

\begin{table}[h]
\begin{center}
    \begin{tabular}{l}
    \hline
   {\bf Algorithm 1. The LE-$\xi$ Algorithm $\hspace{6cm}$}\\ \hline
   \textbf{Hardwire requirement}:\\
   $\hspace{1cm}$$\{\mP_i\}_i$: a collection of processors indexed by node ID $i\in V$;\\
   
   \textbf{Input}:\\
   $\hspace{1cm}$$\{r_j\}_{j\in E}$ with distributed storage where $\mP_i$ keeps $\{r_j\}_{j\in C_i}$;\\
   \textbf{Output}:\\
   $\hspace{1cm}$ $\hat{\theta}=\{\hat{\theta}_j\}_{j\in E}$.\\
   \\
   \textbf{Procedure}:\\
   $\hspace{1cm}$ Operate in parallel for IDs of all non-leaf nodes $\{i\in V:i\notin R\}$\\
   $\hspace{1.5cm}$ If $i\in\sS$,\\
   $\hspace{2cm}$ Get $\hat\xi_j=1-r_j$ for the unique child link $j$ of node $i$ with $\mP_i$;\\
   $\hspace{1.5cm}$ If $i\notin\sS$,\\
   $\hspace{2cm}$ Get $\hat x_i$ with $\mP_i$ from $\{r_j\}_{j\in C_i}$ by solving equation below about $x$\\
   $\hspace{4cm}$ $x=\prod_{j\in C_i}\big[(1-r_j)+r_jx\big]$,\\
   $\hspace{2cm}$ Get $\hat\xi_j=(1-r_j)+r_j\hat x_i$ for every $j\in C_i$ with $\mP_i$;\\
   $\hspace{1cm}$ End parallel operation\\
   \\
   $\hspace{1cm}$ Return $\hat\theta=\Gamma^{-1}(\hat\xi)$.\\ \hline
   \end{tabular}
\end{center}
\end{table}

The validity of the LE-$\xi$ algorithm is guaranteed by the following theorem:
\begin{thm}\label{TheoremLE_xi}
When $n_k\rightarrow\infty$ for all $1\leq k\leq K$, with probability one, the LE-$\xi$ algorithm has a unique solution in $\Theta=(0,1)^m$ that is the MLE.
\end{thm}

\emph{Proof of Theorem \ref{TheoremLE_xi}.}
First, we will show that under the following regularity conditions:
\begin{equation}\label{RegularityCondition}
n_i(1)>0,\ n_i(0)>0\ \ \mbox{and}\ \ \sum_{j\in F_i}n_j(1)<\sum_{j\in B_i}n_j(1)\ \ \mbox{for}\ \forall\ i\in E,
\end{equation}
the likelihood equation with $\xi$ as parameters has a unique solution in $(0,1)^m$.

Note that when (\ref{RegularityCondition}) holds, we have
$$0<r_i<1\ \ \mbox{and}\ \ \sum_{j\in B_i}r_j>1\ \mbox{for}\ \forall\ i\in E.$$
Because the Lemma 1 in \cite{CDHT99} has shown that:
if $0<c_j<1$ and $\sum_jc_j>1$, equation for $x$
$$x=\prod_j\big[(1-c_j)+c_jx\big]$$
has a unique solution in $(0,1)$. It is transparent that under the regularity conditions in (\ref{RegularityCondition}), equation (\ref{LikelihoodEquationPi}) has a unique solution $\hat\pi_i\in(0,1)$ for every non-root link $i$. Considering that for every link $i\in E$,
$$\xi_i=(1-r_i)+r_i\cdot\pi_i\cdot I(i\notin\sS),$$
it is straightforward to see that the likelihood equation with $\xi$ as parameters has a unique solution $\hat\xi\in(0,1)^m$.

Moreover, if
\begin{equation}\label{exception}
\hat\xi\in \Xi,
\end{equation}
we have $\hat\psi=\Lambda(\hat\xi)\in\Psi$. Thus, based on Proposition \ref{LinkTheoremAmongParameters} and Proposition \ref{SuffStat}, we know that if both (\ref{RegularityCondition}) and (\ref{exception}) are satisfied, $\hat\psi$, which is the solution of the likelihood equation with $\psi$ as parameter, is the MLE. Considering the bijections among $\theta$, $\xi$ and $\psi$, this also means that $\hat\theta=\Gamma^{-1}(\hat\xi)$ is the MLE.

Note that the probability that (\ref{RegularityCondition}) or (\ref{exception}) fails goes to zero when $n_k\rightarrow\infty$ for $k=1,\cdots,K$, we complete the proof.\\

The large sample properties of MLE have been well studied by \cite{CDHT99} for tree topology, where the asymptotic normality, asymptotic variance and confidence interval are established. Considering that the likelihood function keeps exactly the same form for both tree and general topology, it is natural to extend these theoretical results for MLE established in \cite{CDHT99} to a network of general topology.\\

With a finite sample, there is a chance that (\ref{RegularityCondition}) or (\ref{exception}) fails. In this case, the MLE of $\theta$ falls out of $\Theta$, and may be missed by the LE-$\xi$ algorithm. For example,
\begin{enumerate}
\item If $n_i(1)=0$ and $n_i(0)>0$, we have $\hat\xi_i=1$;
\item If $n_i(1)>0$ and $n_i(0)=0$, we have $\hat\xi_i=0$;
\item If $\sum_{j\in F_i}n_j(1)=\sum_{j\in B_i}n_j(1)>0$, we have $\hat\theta_i=0$;
\item If $\hat\xi_i\leq\hat\pi_i=\prod_{j\in B_i}\hat\xi_j$ (i.e., $\hat\xi\notin\Xi$), we have $\hat\theta_i\leq0$.
\end{enumerate}
Moreover, if $n_i(1)=n_i(0)=0$, the parameters in subtree $T_i$ (i.e., $\{\theta_j\}_{j\in T_i}$) are not estimable due to lack of information.

In all these cases, we cannot guarantee that the estimate from the LE-$\xi$ algorithm $\hat\theta$ is the global maximum in $\Theta^*=[0,1]^m$. In practice, we can increase sample size by sending additional probes to avoid this dilemma. In case that it is not realistic to send additional probes, we can simply replace $\hat\theta$ by the point in $\Theta^*$ that is the closest to $\hat\theta$, or search the boundary of $\Theta^*$ to maximize $L(\theta)$.

\section{Impacts on the EM Algorithm}\label{sec:NewEM}
The above theoretical results have two major impacts to the EM algorithm widely used in loss tomography. First, as we have shown in Theorem \ref{TheoremLE_xi}, as long as regularity conditions (\ref{RegularityCondition}) and (\ref{exception}) hold, likelihood function $L(\theta)$ has a unique mode in $\Theta$. In this case, the EM algorithm always converges to the MLE for any initial value $\theta^{(0)}\in\Theta$. Considering that the chance of violating (\ref{RegularityCondition}) or (\ref{exception}) goes to zero with the increase of sample size, we have the following corollary for the EM algorithm immediately:

\begin{corollary}\label{CorollaryForEM}
When $n_k\rightarrow\infty$ for all $1\leq k\leq K$, the EM algorithm converges to the MLE with probability one for any initial value $\theta^{(0)}\in\Theta$.
\end{corollary}

Second, the formulation of the new statistics $\{n_i(1)\}_{i\in
E}$ naturally leads to a ``pattern-collapsed" implementation of the EM algorithm, which is computationally much more efficient than the widely used naive implementation where the samples are processed one by one separately. In the naive implementation of the EM algorithm, one enumerates all possible configurations of the internal links compatible with each sample carrying the information on whether the corresponding probe reached the leaf nodes or not. The complexity of the naive implementation in each E-step can be $O(n2^m)$ in the worst case. With the pattern-collapsed implementation, however, the complexity of an E-step can be dramatically reduced to $O(m)$.

The first pattern collapsed EM algorithm is proposed by Deng et al. \cite{Deng2012} in the context of delay tomography. The basic idea is to reorganize the observed data at receivers in delay tomography into \emph{delay patterns}, and make use of a delay pattern database to greatly reduce the computational cost in the E-step. We will show below that the spirit of this method can be naturally extended to the study of loss tomography.

Define event $\{Y_i=0\mid Y_{f_i}=1\}$, or equivalently $\{X_j=0,\
\forall\ j\in R_i\mid X_{f_i}=1\}$, as the \emph{loss event} in
subtree $T_i$, denoted as $\cL_i$. Let $L(\textbf{X}_V;\theta)$ be
the log-likelihood of the complete data where the status of probes
at internal nodes are also observed, $\theta^{(r)}$ be the
estimation obtained at the $r$-th iteration,  the objective function
to be maximized in the $(r+1)$-th iteration of the EM algorithm is:
\begin{eqnarray}\label{Q-function4EM}
Q(\theta,\theta^{(r)})&=& E_{\theta^{(r)}}\big(L(\textbf{X}_V;\theta)\mid\textbf{X}_R\big)=\sum_{t=1}^nE_{\theta^{(r)}}\big(L(X_E^{(t)};\theta)\mid X_R^{(t)}\big)\nonumber\\
&=&\sum_{i\in
E}\Big[n_i(1)\log(1-\theta_i)+n_i(0)Q(\theta_{T_i},\theta^{(r)})\Big],
\end{eqnarray}
where $Q_{\cL_i}(\theta_{T_i},\theta^{(r)})$, which is called the
\emph{localized Q-function} of loss event $\cL_i$, is defined as:
$$Q_{\cL_i}(\theta_{T_i},\theta^{(r)})=E_{\theta^{(r)}}\Big[logP(X_{T_i};\theta_{T_i})\mid\cL_i\Big].$$
Similar to the results for delay patterns shown in the Proposition 1
of \cite{Deng2012}, $Q_{\cL_i}(\theta_{T_i},\theta^{(r)})$ has the
following decomposition:
\begin{equation}\label{QL_decomposition}
Q_{\cL_i}(\theta_{T_i},\theta^{(r)})=\big(1-\psi_i(\theta^{(r)})\big)\log(\theta_i)+\psi_i(\theta^{(r)})\Big[\log(1-\theta_i)+\sum_{j\in
C_i}Q_{\cL_j}(\theta_{T_j},\theta^{(r)})\Big].
\end{equation}
Integrating (\ref{Q-function4EM}) and (\ref{QL_decomposition}), we
have
$$Q(\theta,\theta^{(r)})=\sum_{i\in E}\Big[\omega_i(1)\log(1-\theta_i)+\omega_i(0)\log(\theta_i)\Big],$$
where $\{\omega_i(1),\omega_i(0)\}_{i\in E}$ are defined recursively
as follows:

\noindent$\hspace{0.5cm}\bullet$ for $i\in\{S_1,\cdots,S_K\}$, i.e., being one of the root links,
\begin{eqnarray*}
\omega_i(1)&=&n_i(1)+n_i(0)\cdot\psi_i(\theta^{(r)}),\\
\omega_i(0)&=&n_i(0)\cdot\big(1-\psi_i(\theta^{(r)})\big);
\end{eqnarray*}
$\hspace{0.5cm}\bullet$ for $i\notin\{S_1,\cdots,S_K\}$,
\begin{eqnarray*}
\omega_i(1)&=&n_i(1)+\sum_{j\in F_i}\omega_j(0)\cdot\psi_j(\theta^{(r)}),\\
\omega_i(0)&=&\sum_{j\in
F_i}\omega_j(0)\cdot\big(1-\psi_j(\theta^{(r)})\big).
\end{eqnarray*}
And, the M-step is:
$$\theta_i^{(r+1)}=\frac{\omega_i(0)}{\omega_i(0)+\omega_i(1)},\ \forall\ i\in E.$$

In this paper, we refer to the EM with the pattern-collapsed implementation as PCEM to distinguish from the EM with the naive implementation, which is referred to as NEM. We have shown in \cite{Deng2012} by simulation and theoretical analysis that PCEM is computationally much more efficient than the naive EM in the context of delay tomography. In context of loss tomography, it is easy to see from the above analysis that PCEM enjoys a complexity of $O(m)$ for each E-step, which is much more efficient than the naive EM, whose complexity can be $O(n2^m)$ for each E-step in the worst case.

\section{Simulation Studies}\label{sec:Simulation}
We verify the following facts via simulations:
\begin{enumerate}
\item PCEM obtains exactly same results as the naive EM for both tree topology and general topology, but is much faster;
\item LE-$\xi$ obtains exactly same results as the EM algorithm when regularity conditions (\ref{RegularityCondition}) and (\ref{exception}) hold, but is much faster when the speedup from parallel computation is considered.
\end{enumerate}


We carried out simulations on a 5-layer network covered by two trees as showed in Figure \ref{L5net}.
The network has 49 nodes labeled from Node 0 to Node 48, and two sources Node 0 and Node 32. 
We conduct two set of simulations. One is based on the ideal model, and the other is by using network simulator 2 (ns-2, \cite{NS2}).

\subsection{Simulation study with the ideal model}
In the first set of simulations, the data is generated from the ideal model where each link has a pre-defined constant loss rate.

To test the performance of methods on different magnitude of loss rates, we draw the link-level loss rates from 
Beta distributions with different parameters.  
The link-level loss rates $\theta=\{\theta_i\}_i$ are randomly sampled from $\text{Beta}(1,99)$, $\text{Beta}(5,995)$, 
$\text{Beta}(2,998)$ or $\text{Beta}(1,999)$.  For example, the mean of $\text{Beta}(1,999)$ is $0.001$, so if we draw loss rates from
$\text{Beta}(1,999)$, then on average we expect a $0.1\%$ link-level loss rate in the network.
Given the loss rate vector $\theta$ for each network, we generated $100$ independent datasets with sample sizes $50$, $100$, $200$ and $500$, respectively.
Thus, a total of $100\times4\times4=1600$ datasets were simulated. The sample size is evenly distributed into the two trees, e.g., the source of each tree will send out $100$ probes when sample size $n=200$.

To each of the 1600 simulated data sets, we applied NEM, PCEM, LE-$\xi$ and MVWA, respectively. The stopping rule of the EM algorithms is 
$\max_i|\theta_i^{(t+1)}-\theta_i^{(t)}|\leq10^{-6}$, the initial values are $\theta_i^{(0)}=0.03$ for all $\ i\in E$. 
We implement MVWA algorithm by obtaining MLE estimate in each tree with LE-$\xi$ algorithm. Table \ref{MSE_L5} 
summarize the average running time and MSEs of these methods under different settings. From the tables, we can see that:

\begin{enumerate}
\item The performances of all four methods in term of MSE improve with the increase of sample size $n$ in both networks.

\item NEM and PCEM always get exactly same results.

\item The result from LE-$\xi$ is slightly different from these of the EM algorithms when sample size $n$ is as small as $100$ or $200$ due to the violation of the regularity conditions (\ref{RegularityCondition}) and (\ref{exception}). 
When $n=500$, the results of LE-$\xi$ and EM algorithms become identical.

\item LE-$\xi$, PCEM and MVWA algorithm implemented with LE-$\xi$ are dramatically faster than NEM. For example, in simulation configuration with $\text{Beta}(1,1000)$ and sample size $n=500$, the
running time of NEM is almost 2,000 times of LE-$\xi$ and PCEM.

\item The MSE of MVWA is always larger than the MSE of the other three methods, which capture the MLE. This is consistent with the result in \cite{BDPT}.
\end{enumerate}


\begin{table}
\caption{Performance of different methods on simulated data from ideal model on the 5-layer network in Figure \ref{L5net}.} \label{MSE_L5}\setlength{\tabcolsep}{1.5pt}{\tiny
\begin{center}

\begin{tabular}{cccccccccc}
\hline 
 &  & \multicolumn{2}{c}{Beta(1,100)} & \multicolumn{2}{c}{Beta(5,1000)} & \multicolumn{2}{c}{Beta(2,1000)} & \multicolumn{2}{c}{Beta(1,1000)}\tabularnewline
$n$ & Method & Time (ms) & MSE (1e-5) & Time (ms) & MSE (1e-5) & Time (ms) & MSE (1e-5) & Time (ms) & MSE (1e-5)\tabularnewline
\hline 
\multirow{4}{*}{50} & NEM & 5065.50  & 764.49  & 4444.10  & 364.88  & 3396.20  & 188.78  & 2778.15  & 91.10 \tabularnewline
 & PCEM & 2.25  & 764.49  & 2.55  & 364.88  & 2.55  & 188.78  & 2.60  & 91.10 \tabularnewline
 & LE-$\xi$ & 2.15  & 768.14  & 2.25  & 365.44  & 2.15  & 189.12  & 2.20  & 91.22 \tabularnewline
 & MVWA & 2.35  & 772.01  & 2.05  & 368.78  & 2.35  & 189.70  & 2.60  & 91.34 \tabularnewline
\cline{1-10} 
\multirow{4}{*}{100} & NEM & 12102.90  & 434.39  & 9109.65  & 245.08  & 7323.90  & 89.64  & 7112.55  & 36.69 \tabularnewline
 & PCEM & 3.85  & 434.39  & 3.90  & 245.08  & 3.90  & 89.64  & 4.85  & 36.69 \tabularnewline
 & LE-$\xi$ & 3.70  & 436.41  & 3.50  & 247.29  & 3.80  & 90.02  & 4.30  & 36.74 \tabularnewline
 & MVWA & 4.70  & 439.50  & 4.40  & 247.36  & 4.90  & 90.08  & 5.00  & 36.80 \tabularnewline
\cline{1-10} 
\multirow{4}{*}{200} & NEM & 31368.95  & 212.79  & 23558.50  & 109.03  & 19184.65  & 44.79  & 16249.45  & 18.52 \tabularnewline
 & PCEM & 12.10  & 212.79  & 11.05  & 109.03  & 10.20  & 44.79  & 11.20  & 18.52 \tabularnewline
 & LE-$\xi$ & 11.65  & 212.87  & 10.65  & 109.69  & 10.70  & 44.91  & 10.75  & 18.54 \tabularnewline
 & MVWA & 13.90  & 213.90  & 11.05  & 109.75  & 11.80  & 44.90  & 10.20  & 18.55 \tabularnewline
\cline{1-10} 
\multirow{4}{*}{500} & NEM & 90221.95  & 85.17  & 49625.15  & 43.88  & 45402.80  & 14.35  & 43475.15  & 6.98 \tabularnewline
 & PCEM & 25.05  & 85.17  & 22.25  & 43.88  & 22.70  & 14.35  & 22.35  & 6.98 \tabularnewline
 & LE-$\xi$ & 28.05  & 85.17  & 21.75  & 43.88  & 21.30  & 14.35  & 23.10  & 6.98 \tabularnewline
 & MVWA & 27.00  & 85.45  & 22.50  & 44.02  & 28.10  & 14.38  & 24.25  & 6.98 \tabularnewline
\hline 
\end{tabular}

\end{center}}
\end{table}



\subsection{Simulation study by network simulator 2}

We also conduct the simulation study using ns-2. 
We use the network topology shown in Figure \ref{L5net}, where the two sources located at Node 0 and Node 32 
multicast probes to the attached receivers. Besides the network traffic created by multicast probing, 
a number of TCP sources with various window sizes and a number of UDP sources with different burst rates and 
periods are added at several nodes to produce cross-traffic. The TCP and UDP cross-traffic takes about $80\%$ 
of the total network traffic. We generated $100$ independent datasets by running the ns-2 simulation for 1, 2, 
5 or 10 simulation seconds. The longer experiments generate more multicast probes samples. We record all pass 
and loss events for each link during the simulation, and the actual loss rates are calculated and considered 
as the true loss rates for calculating MSEs. Table \ref{MSE_ns} shows the average number of 
packets (sample size), running time of methods and MSEs for different ns-2 simulation times.

In Table \ref{MSE_ns}, we observe similar results as shown in Table \ref{MSE_L5}. The performances of all four 
methods in term of MSE improve with the increase of sample size $n$ in both networks. The result from LE-$\xi$ is 
slightly different from these of the EM algorithms when sample size $n$ is as small. More importantly, LE-$\xi$ and 
PCEM are dramatically faster than NEM.

\begin{table}
\caption{Performance of different methods on simulated data from network simulator 2.} \label{MSE_ns}\setlength{\tabcolsep}{3pt}{\tiny
\begin{center}
\begin{tabular}{cccc}
\hline 
Sim time ($n$) & Method & Time (ms) & MSE (1e-5)\tabularnewline
\hline 
\multirow{4}{*}{1s (528)} & NEM & 54675.20  & 581.29 \tabularnewline
 & PCEM & 21.80  & 581.29 \tabularnewline
 & LE-$\xi$ & 22.00  & 581.29 \tabularnewline
 & MVWA & 22.80  & 584.50 \tabularnewline
\cline{1-4} 
\multirow{4}{*}{2s (2038)} & NEM & 256270.60  & 232.49 \tabularnewline
 & PCEM & 21.80  & 232.49 \tabularnewline
 & LE-$\xi$ & 22.00  & 232.49 \tabularnewline
 & MVWA & 22.80  & 233.63 \tabularnewline
\cline{1-4} 
\multirow{4}{*}{5s (5648)} & NEM & 745393.40  & 122.23 \tabularnewline
 & PCEM & 329.40  & 122.23 \tabularnewline
 & LE-$\xi$ & 315.00  & 122.23 \tabularnewline
 & MVWA & 299.10  & 122.86 \tabularnewline
\cline{1-4} 
\multirow{4}{*}{10s (13355)} & NEM & 1334478.90  & 75.24 \tabularnewline
 & PCEM & 612.90  & 75.24 \tabularnewline
 & LE-$\xi$ & 612.10  & 75.24 \tabularnewline
 & MVWA & 657.00  & 76.59 \tabularnewline
\cline{1-4} 
\end{tabular}
\end{center}}
\end{table}


\section{Conclusion}\label{sec:Conclusion}
We proposed a set of sufficient statistics called \emph{internal view} and two alternative parameter systems $\xi$ and $\psi$ for loss tomography. We found that the likelihood function keeps the exactly same formulation for both tree and general topologies under all three types of parametrization ($\theta$, $\xi$ and $\psi$), and we can switch among the three parameter systems freely without changing the result of parameter estimation. We also discovered that the parameterization of the likelihood function based on $\psi$ falls into the standard exponential family, which has a unique mode in parameter space $\Psi$ under regularity conditions, and the parametrization based on $\xi$ leads to an efficient algorithm called LE-$\xi$ to calculate the MLE, which can be carried out in a parallel fashion. These results indicate that loss tomography for general topologies enjoys the same mathematical nature as that for tree topologies, and can be resolved effectively. The proposed statistics and alternative parameter systems also lead to a more efficient pattern-collapsed implementation of the EM algorithm for finding MLE, and a theoretical promise that the EM algorithm converges to the MLE with probability one when sample size is large enough. Simulation studies confirmed our theoretical analysis as well as the superiority of the proposed methods over existing methods.


\appendix

\end{document}